\documentclass{amsart}
\usepackage{amsfonts}

\theoremstyle{plain}

\numberwithin{equation}{section}

\input{tcilatex}

\begin{document}
\title[Quantum noise and output process]{The world of quantum noise and the
fundamental output process}
\author{V. P. Belavkin}
\address[A. One and A. Three]{ Mathematics Department\\
University of Nottingham, UK}
\email{vpb@maths.nott.ac.uk}
\urladdr{http://www.maths.nott.ac.uk/personal/vpb/}
\thanks{The first author thanks to Matt James for support and hospitality in
ANU, Canberra, where this paper was prepared for the current publication }
\author{O. Hirota}
\address{Research Center of Quantum Communications\\
Tamagawa University, Tokyo, Japan}
\email{}
\urladdr{}
\thanks{}
\author{R. Hudson}
\address{}
\urladdr{}
\date{September 26, 1994}
\dedicatory{}
\thanks{This paper was originally published in: Proceedings of the 2nd QCMC
conference: \textit{Quantum Communications and Measurement,} 3--19, Plenum
Press, New-York and London 1995.}

\begin{abstract}
A stationary theory of quantum stochastic processes of second order is
outlined. It includes KMS processes in wide sense like the equilibrium
finite temperature quantum noise given by the Planck's spectral formula. It
is shown that for each stationary noise there exists a natural output process%
\index{output process} which is identical to the noise in the infinite
temperature limit, and flipping with the noise if the time is reversed at
finite temperature. A canonical Hilbert space representation of the quantum
noise and the fundamental output process is established and a decomposition
of their spectra is found. A brief explanation of quantum stochastic
integration with respect to the input-output processes is given using only
correlation functions. This provides a mathematical foundation for linear
stationary filtering transformations of quantum stochastic processes. It is
proved that the colored quantum stationary noise and its time-reversed
version can be obtained in the second order theory by a linear nonadapted
filtering of the standard vacuum noise uniquely defined by the canonical
creation and annihilation operators on the spectrum of the input-output pair.
\end{abstract}

\maketitle

\section{Introduction}

In this paper we develop a correlation theory of stationary quantum noise,
give its spectral analysis%
\index{spectral analysis} and classification and extend the theory of
quantum stochastic integration%
\index{stochastic integration|(} \cite{bib:noise1} to colored quantum noise 
\cite{bib:noise2}. The typical example of such noise is given by the Planck
spectral formula. It shows that finite temperature equilibrium quantum noise
is not white in contrast to classical equilibrium noise given by the Nyquist
spectral formula. The weak coupling limit and rotating wave approximation 
\cite{bib:noise2,bib:noise3} make it possible to approximate the equilibrium
quantum noise by canonical pairs of noncommutative white noises in a narrow
spectral band near a resonant frequency.

Although such an approximation is sufficient in quantum optics%
\index{quantum optics} and in many other practical cases, it is not
satisfactory from a purely theoretical point of view because it does not
correctly predict the simplest quantum stochastic motion of a free Brownian
particle in an equilibrium environment, when the resonant frequency is zero.

In order to describe the output stochastic processes of quantum filters and
other devices of quantum measurement%
\index{quantum measurement} and communications as functional transformations
of the equilibrium noise, we also need a generalization of the notion of
quantum output fields \cite{bib:noise14,bib:noise4,bib:noise5} in the
framework of stochastic integration with respect to the colored quantum noise%
\index{quantum noise}

The well developed quantum stochastic calculus and noncommutative extension
of It\^{o} integration to vacuum noise with canonical commutation relations
in the time domain \cite{bib:noise5,bib:noise6} is not directly applicable
to such cases, neither is the calculus and integration with respect to
stochastic models of temperature quantum noise with flat spectrum \cite%
{bib:noise4,bib:noise7}.

But we can use the methods of quantum stochastic integration in the spectral
domain, in which the stationary quantum noise is $\delta $-correlated with a
certain modification. The latter is required because the colored quantum
processes are not frequency-stationary and need not be adapted in the
spectral representation. Although non-stationary and non-adapted theories of
quantum stochastic calculus and integration have been already established 
\cite{bib:noise9,bib:noise8,bib:noise10}, here we will use the much more
simple approach of mean square integration which is sufficient for the
non-adapted linear transformations. The corresponding classical theory of
stochastic integration was developed by Wiener before the It\^{o}
integration theory and is based on the possibility of representing any
(colored) process stationary in the wide sense as a linear filtering%
\index{linear filtering} integration of the standard white noise.

Here we shall prove that any quantum process stationary in the wide sense
can be also obtained from a standard one also by filtering. In fact, we will
show that it can be obtained by mean square integration with respect to a
canonical pair of orthogonal quantum integrators describing the standard
zero temperature (vacuum) noise in the second order. Although such a
possibility is known for the Gaussian case as a Bogolubov transformation
which doubles degrees of freedom of the noise in such representation, it has
not been realized in the theory of quantum noise and integration.

We will remove this unpleasant feature by deriving the fundamental output
process for a given quantum noise as the canonically time-reversed noise,
which commutes with the input noise and has maximal possible correlations
with it. In the classical (or infinite temperature) case, the fundamental
output process coincides with the noise. In the quantum case it gives the
best possible nondemolition filtering and time continuous indirect
observation of the noise.

The fundamental output process for a given Gaussian quantum noise in the
framework of quantum stochastic calculus was introduced in \cite{bib:noise11}
and the possibility of its nondemolition observation was demonstrated for
Markovian models of interaction with a quantum open system in the quantum
theory of filtering (see the recent survey \cite{bib:noise12} and papers
cited therein).

To explain the idea of the fundamental output process, let us represent a
canonical Bose-noise annihilation and creation pair $\mathrm{b}$, $\mathrm{b}%
^{\dagger }$ with non-zero temperature correlations

\begin{equation*}
\langle \mathrm{b}^{\dag }\mathrm{b}\rangle =n,\ \ \ \langle \mathrm{bb}%
^{\dag }\rangle =n+1
\end{equation*}%
by the linear combination $\mathrm{b}=%
\sqrt{n+1}\mathrm{a}+\sqrt{n}\mathrm{c}$ of the zero temperature pairs $%
\mathrm{a}$, $\mathrm{a}^{\dagger }$ and $\mathrm{c}$, $\mathrm{c}^{\dagger
} $ with

\begin{equation*}
\lbrack \mathrm{a},\mathrm{a}^{\dag }]=1,\ \ [\mathrm{c}^{\dag },\mathrm{c}%
]=1,\ \ [\mathrm{a}^{\dag },\mathrm{c}]=[\mathrm{a},\mathrm{c}]=0.
\end{equation*}%
Then the combination $\mathrm{\tilde{b}}=\sqrt{n}\mathrm{a}+\sqrt{n+1}%
\mathrm{c}$ defines a canonical pair $\mathrm{\tilde{b}}$, $\mathrm{\tilde{b}%
}^{\dagger }$ of output creation and annihilation operators with correlations

\begin{equation*}
\langle \mathrm{\tilde{b}}^{\dag }\mathrm{\tilde{b}}\rangle =n+1,\ \ \langle 
\mathrm{\tilde{b}\tilde{b}}^{\dag }\rangle =n
\end{equation*}%
which commutes with the noise, $[\mathrm{\tilde{b}},\mathrm{b}]=[\mathrm{%
\tilde{b}},\mathrm{b}^{\dag }]=0$, and provides the maximal mutual
correlations $\langle \mathrm{\tilde{b}b}^{\dag }\rangle =\sqrt{n(n+1)}$,
corresponding to correlation coefficient $r=1$.

In the classical case, the latter would mean that $\mathrm{\tilde{b}}=%
\mathrm{b}$, but this is not the case: $\mathrm{\tilde{b}}$ commutes with $%
\mathrm{b}^{\dag }$ but $\mathrm{b}$ does not. Nevertheless $\mathrm{\tilde{b%
}}$ is uniquely defined for the noise $\mathrm{b}$ by the described
properties as the time-reversed noise, which in the real arrow of time plays
the role of the fundamental output process. Moreover, the zero temperature
pair $\mathrm{a}$, $\mathrm{c}$ is uniquely determined by the pair $(\mathrm{%
b},\mathrm{\tilde{b}})$, being given by

\begin{equation*}
\mathrm{a}=\sqrt{n+1}\mathrm{b}-\sqrt{n}\mathrm{\tilde{b}},\ \ \ \mathrm{c}=%
\sqrt{n+1}\mathrm{\tilde{b}}-\sqrt{n}\mathrm{b}.
\end{equation*}


\section{Quantum Correlations%
\index{quantum correlation} and Re\-ve\-r\-s\-ed Pro\-ce\-ss\-es}

Here we sketch of a \emph{second order theory} of quantum noise and linear
filtering of stationary signals, initiated more than 20 years ago in the
pioneer paper \cite{bib:noise13}. A real scalar quantum stochastic noise as
a process $\mathrm{x}_{\cdot }=\left( \mathrm{x}_{j}\right) $ in second
order is completely determined by the zero mean values $\langle \mathrm{x}%
_{j}\rangle =0$ and finite variances alone $\langle \mathrm{x}%
_{j}^{2}\rangle <\infty $ of the Hermitian variables $\mathrm{x}_{j}=\mathrm{%
x}_{j}^{\dagger }$ and also by their not necessarily symmetric covariances,
or correlations $\langle \mathrm{x}_{i}\mathrm{x}_{j}$$\rangle \neq \langle 
\mathrm{x}_{j}\mathrm{x}_{i}\rangle $. The products $\mathrm{x}_{i}\mathrm{x}%
_{j}$ in $\langle \cdot \rangle $ form a Hermitian-positive kernels $%
\left\langle \mathrm{x}_{i}|\mathrm{x}_{j}\right\rangle $ which define the
scalar products in a complex Euclidean space. By writing an invertible
correlation kernel $\langle \mathrm{x}_{i}\mathrm{x}_{j}$$\rangle \equiv
\langle \mathrm{x}_{i}|\mathrm{x}_{j}$$\rangle $ as a bra-ket inner product $%
\langle \mathrm{x}_{i}|\cdot |\mathrm{x}_{j}$$\rangle $, we can describe the
quantum noise in second order by real vectors $|\mathrm{x}_{j}\rangle =|%
\mathrm{x}_{j}\rangle ^{\sharp }$ equipped with the involution $|\mathrm{z}%
\rangle \mapsto |\mathrm{z}\rangle ^{\sharp }=|\mathrm{z}^{\dagger }\rangle $%
. The latter, defined by%
\begin{equation*}
|{{\sum\limits_{{}}^{{}}\zeta _{j}{\mathrm{x}}_{j}}}\rangle ^{\sharp
}=\sum\limits_{{}}^{{}}\zeta _{j}^{\ast }|{\mathrm{x}}_{j}\rangle \;\;\;\;\
\ \ \forall \zeta _{j}\in \mathbf{C}\,
\end{equation*}%
such that$\ {|}${$\mathrm{z}$}${\rangle }^{\sharp \sharp }={|}${$\mathrm{z}$}%
${\rangle }$, gives vector representation of the Hermitian conjugation $%
\mathrm{z}\mapsto \mathrm{z}^{\dag }=\sum\limits_{{}}^{{}}{\zeta }_{j}^{\ast
}\mathrm{x}_{j}$ with $\zeta _{j}^{\ast }=%
\overline{\zeta _{j}}$ for the complex combinations $\mathrm{z}%
=\sum\limits_{{}}^{{}}\zeta _{j}\mathrm{x}_{j}\equiv \zeta \cdot \mathrm{x}$
of the Hermitian variables $\mathrm{x}{_{j}}$ given by raws $\zeta _{\cdot
}=\left( \zeta _{i}\right) $.

Let us interpret the index $j$ as \textquotedblleft discrete
time\textquotedblright , e.g. $j\in \mathbf{Z}$\ when $\mathrm{x}_{\cdot }=(%
\mathrm{x}{_{j})}_{j\in \mathbf{Z}}$ is two sided infinite sequence $(\cdots
,\mathrm{x}_{-1},\mathrm{x}_{0},\mathrm{x}_{1},\cdots {)}$ of Hermitian
variables with the time ordered correlations 
\begin{equation*}
K_{ij}=\left\langle \mathrm{x}_{i}\ \mathrm{x}_{j}\right\rangle \equiv
\left\langle \mathrm{x}_{i}|\mathrm{x}_{j}\right\rangle ,\ \ \ i>j\in 
\mathbf{Z}
\end{equation*}%
and the reverse ordered $K_{ji}=\left\langle \mathrm{x}_{j}|\mathrm{x}%
_{i}\right\rangle $ which we need to define the scalar product of the
corresponding vectors $|${$\mathrm{x}$}$_{j}\rangle $. In contrary to the
classical theory, these correlations need not be real valued even for
Hermitian $\mathrm{x}_{j}$, so that time reversal corresponds to their
complex conjugation: 
\begin{equation*}
K_{ji}=\left\langle \mathrm{x}_{j}\ \mathrm{x}_{i}\right\rangle =\overline{%
\left\langle \ \mathrm{x}_{i}\ \mathrm{x}_{j}\right\rangle }=\overline{K}%
_{ij}.
\end{equation*}%
Although the reversed correlations may be considered as not
\textquotedblleft observable\textquotedblright\ in the real arrow of time,
they are defined by complex conjugation in such a way that the Hermitian
matrix $K=[{K_{ij}]}$ is positive definite: 
\begin{equation*}
{\zeta }^{\ast }K\cdot \zeta :=\sum\limits_{{}}^{{}}{\zeta }_{i}^{\ast
}\left\langle \mathrm{x}_{i}\ \mathrm{x}_{j}\right\rangle {\zeta }%
_{j}=\left\langle \mathrm{z}^{\dag }\ \mathrm{z}\right\rangle \geq 0,
\end{equation*}%
as is the complex conjugate = transpose $\overline{K}=[{\overline{K}}%
_{ij}]=[K_{ji}]=\widetilde{K}:$ 
\begin{equation*}
{\zeta }^{\ast }\widetilde{K}\cdot \zeta :=\sum\limits_{{}}^{{}}{\zeta }%
_{j}\left\langle \mathrm{x}_{j}\ \mathrm{x}_{i}\right\rangle {\zeta }%
_{i}^{\ast }=\left\langle {\ }\mathrm{z}\ \mathrm{z}{^{\dag }}\right\rangle
\geq 0.
\end{equation*}%
We claim that $\overline{K}$ can be treated as the correlation matrix 
\begin{equation*}
\widetilde{K}_{ij}=\left\langle \widetilde{\mathrm{x}}_{i}\ \widetilde{%
\mathrm{x}}_{j}\right\rangle \equiv \langle \widetilde{\mathrm{x}}_{i}|%
\widetilde{\mathrm{x}}_{j}\rangle ,\ \ \ \forall i\geq j\in \mathbf{Z}
\end{equation*}%
of another sequence ${{{\widetilde{\mathrm{x}}}_{i}}}=(\cdots ,\widetilde{%
\mathrm{x}}_{-1},\widetilde{\mathrm{x}}_{0},\widetilde{\mathrm{x}}%
_{1},\cdots )$ of Hermitian $\widetilde{\mathrm{x}}_{i}$ commuting with $%
\mathrm{x}_{j}$ for all $j\in \mathbf{Z}$ which is maximally correlated to $%
\mathrm{x}_{\cdot }=(\mathrm{x}_{j})_{j\in \mathbf{Z}}$ in the sense that
the real covariance matrix $G=\left[ \left\langle \widetilde{\mathrm{x}}%
_{i}\ \mathrm{x}_{j}\right\rangle \right] $ ispositive and symmetric as the
geometric mean $G=\left( K\cdot \widetilde{K}\right) ^{1/2}$ of $K$ and $%
\widetilde{K}=\overline{K}$. The sequence $\widetilde{\mathrm{x}}$ describes
the fundamental output process as reversed noise by the quantum Hermitian
variables $\widetilde{\mathrm{x}}_{j}={\ \widetilde{\mathrm{x}}}_{j}^{\dag }$
which will also be represented by real vectors $|\widetilde{\mathrm{x}}%
_{j}\rangle =|\widetilde{\mathrm{x}}_{j}\rangle ^{\flat }$ with respect to
another involution ${|\widetilde{\mathrm{z}}\rangle }\mapsto {|\widetilde{%
\mathrm{z}}\rangle }^{\flat }$ uniquely defined on the complex combination ${%
\widetilde{\mathrm{z}}=}\zeta \cdot \widetilde{\mathrm{x}}$ by complex
conjugation of the coefficients $\zeta _{j}\in \mathbf{C}$. This suggests a
natural norm 
\begin{equation*}
\left\Vert \zeta _{\cdot }\right\Vert ^{2}=\left\langle {{\mathrm{z}}^{\dag
}\ \mathrm{z}}\right\rangle +\left\langle {\mathrm{z}\ {\ \mathrm{z}}^{\dag }%
}\right\rangle ={\zeta }^{\ast }\left( {\ K+\widetilde{K}}\right) \cdot \zeta
\end{equation*}%
in the space $\mathcal{E}$ of test sequences $\zeta _{\cdot }=\left( \zeta
_{i}\right) $ for the simultaneous treatment of linear combinations $|%
\mathrm{z}\rangle =\sum\limits_{{}}^{{}}{\zeta }_{j}|\mathrm{x}_{j}\rangle $
of the $\sharp $ - real vectors $|\mathrm{x}_{j}\rangle ={|}\mathrm{x}_{j}{%
\rangle }^{\sharp }$, and the combinations $|{\widetilde{\mathrm{z}}}\rangle
=\sum\limits_{{}}^{{}}{\zeta }_{j}|{\widetilde{\mathrm{x}}}_{j}\rangle $ of $%
\flat $ - real vectors $|\widetilde{\mathrm{x}}_{j}\rangle ={|}\widetilde{%
\mathrm{x}}_{j}{\rangle }^{\flat }$, representing ${\widetilde{\mathrm{x}}}%
_{j}=\widetilde{\mathrm{x}}_{j}^{\dag }$.

From now on for the sake of simplicity we shall restrict ourselves to the
case when the matrices $K$ and $\widetilde{K}$ commute as it is in the
stationary case 
\begin{equation*}
K_{ij}=k_{i-j},\ \ \ k_{-j}=\overline{k}_{j},\ \ \ \forall j\in \mathbf{Z}
\end{equation*}%
(see also \cite{bib:noise11} for a more general, noncommutative case.) In
this case that covariances of {$\mathrm{x}$}$_{i}$ and ${{\widetilde{\mathrm{%
x}}}}_{j}$ are described by the real symmetric matrix 
\begin{equation*}
G={\left( {K\ \overline{K}}\right) }^{1/2}={\left( {\overline{K}\ K}\right) }%
^{1/2}=\overline{G}
\end{equation*}%
as geometric mean of the commuting ${K}$ and ${\overline{K}}$, corresponding
to maximal correlations with zero commutators between $\mathrm{x}$ and $%
\widetilde{\mathrm{x}}$: 
\begin{equation*}
{G}_{ij}=\left\langle {{\mathrm{x}}_{i}\ {\widetilde{\mathrm{x}}}_{j}}%
\right\rangle =\left\langle {{\widetilde{\mathrm{x}}}_{j}\mathrm{x}_{i}}%
\right\rangle =G_{ji},\ \ \ \forall i>j\in \mathbf{Z}
\end{equation*}%
If $K$ is invertible, this output process is identically correlated with the
noise, corresponding to the correlation coefficient $r=1$, and in the
classical case $\widetilde{K}=K$ such the sequence $\left( {{\widetilde{%
\mathrm{x}}}_{j}}\right) $ is always identified with $\left( {{\mathrm{x}}%
_{j}}\right) $. The invertible case corresponds to the thermal noise which
we call \emph{standard} if $G_{ij}=\delta _{ij}$, that is if the inverse
noise $\left( \widetilde{\mathrm{x}}_{j}\right) $ is $\delta $-correlated
with $\left( \mathrm{x}_{j}\right) $ such that $\widetilde{K}={K}^{-1}$. The
opposite situation $\widetilde{K}K=0$ when $\left( \widetilde{\mathrm{x}}%
_{j}\right) $ is uncorrelated with $\left( \mathrm{x}_{j}\right) $ such that 
$r=0$ is possible only in the quantum case. It corresponds to the vacuum
noise which we call also \emph{standard} if $K+\widetilde{K}=I$.

Let us prove the existence of the time-reversed sequence $\widetilde{\mathrm{%
x}}\neq \mathrm{x}$ in the second order quantum theory corresponding to the
case $\widetilde{K}\neq K$, and find the adjoint involution $|$\textrm{$%
\tilde{z}$}$\rangle \mapsto {|}${$\mathrm{\tilde{z}}$}${\rangle }^{\flat }=|%
\mathrm{\tilde{z}}^{\dagger }\rangle $, 
\begin{equation*}
{\ {\ |\sum\limits_{{}}^{{}}\zeta _{j}{\widetilde{\mathrm{x}}}_{j}\rangle }}%
=\sum\limits_{{}}^{{}}\zeta _{j}^{\ast }|{\widetilde{\mathrm{x}}}_{j}\rangle
\ \ \ \ \forall \zeta _{j}\in \mathbf{C}
\end{equation*}%
such that$\ {|}${$\mathrm{\tilde{z}}$}${\rangle }^{\flat \flat }={|}${$%
\mathrm{\tilde{z}}$}${\rangle }$, giving the reversed vector representation
of the Hermitian conjugation $\mathrm{\tilde{z}}\mapsto ${$\mathrm{\tilde{z}}
$}$^{\dag }=\sum \zeta _{j}^{\ast }{\widetilde{\mathrm{x}}}_{j}$ for the
complex combinations $\mathrm{\tilde{z}}=\sum \zeta _{j}{\widetilde{\mathrm{x%
}}}_{j}$ in the case of invertible $K$.

The commuting matrices $K$ and $\widetilde{K}$ have common eigen vectors $%
u_{\cdot }$ (generalized row-vectors), given as complex eigen-sequences $%
u_{\cdot }={\left( {u_{j}}\right) }_{j\in \mathbf{Z}}$ such that 
\begin{equation*}
\sum\limits_{i\in \mathbf{Z}}^{{}}u_{i}K_{ij}={\kappa }u_{j},\ \ \
\sum\limits_{i\in {\ }\mathbf{Z}}^{{}}u_{i}{\widetilde{K}}_{ij}=\widetilde{%
\kappa }u_{j},\ \ \ j\in \mathbf{Z}
\end{equation*}%
with eigenvalues ${\kappa }$ and $\widetilde{\kappa }\geq 0$, corresponding
to the complex conjugate eigen sequences $\overline{u_{\cdot }}={\left( {{%
\overline{u}}_{j}}\right) }_{j\in \mathbf{Z}}$ : 
\begin{equation*}
\sum\limits_{i\in \mathbf{Z}}^{{}}\overline{u}_{i}K_{ij}=\widetilde{\kappa }%
\overline{u}_{j},\ \ \ \sum\limits_{i\in \mathbf{Z}}^{{}}\overline{u}_{i}%
\widetilde{K}_{ij}=\kappa \overline{u}_{j},\ \ \ j\in \mathbf{Z}
\end{equation*}

Indeed, let $K$ and $\widetilde{K}$ have a common eigen-row-vector $u_{\cdot
}$ with eigenvalues $\kappa \geq 0$ and $\widetilde{\kappa }\geq 0$
respectively. Because $\widetilde{K}$ is the complex conjugate matrix of $K$%
, $\overline{u_{\cdot }}$ is also a common eigenvector, but with the real
eigenvalue $\widetilde{\kappa }$ for $K$ and $\kappa $ for $\widetilde{K}$.
Two cases arise

\begin{description}
\item[$\protect\kappa=\bar{\protect\kappa}$] In this case we may assume
without loss of generality that $u_j=\overline{u}_j,\,\forall j\in \mathbf{Z}
$

\item[$\protect\kappa \neq \tilde{\protect\kappa}$] In this case the
eigenvectors $u_{\cdot }$, $\overline{u_{\cdot }}$ are orthogonal and so we
have $\sum u_{j}^{2}=0$.
\end{description}

This means that the index set $\Omega $ of the eigen sequences $\langle \nu
|:=(u_{j}(\nu ))_{j\in \mathbf{Z}}$ is equipped with a natural flip 
\begin{equation*}
\nu \mapsto -\nu ,\ \ \ -(-\nu )=\nu ,\ \ \ \widetilde{\kappa }(\nu )=\kappa
(-\nu )
\end{equation*}%
such that $\overline{u_{\cdot }}(\nu )=u_{\cdot }(-\nu )$ and the
(generalized) Plansherel measure $\mathrm{d}\nu $ on $\Omega $ is invariant
under this flip. The complete (with respect to $\mathrm{d}\nu $) orthogonal
set $\left\{ u_{\cdot }{(\nu )|\nu \in \Omega }\right\} $ can be chosen in
such a way that $u_{j}(\nu )=u_{j}(-\nu )$ iff ${\kappa (\nu )}=\kappa (-\nu
)$ and $\sum\limits_{j\in \mathbf{Z}}^{{}}u_{j}^{2}(\nu )=0$ iff $\kappa
(\nu )\neq \kappa (-\nu )$. This flip defined an isometric involution $%
\varphi ^{\star }\left( \nu \right) :=\overline{\varphi }\left( -\nu \right)
\equiv \widetilde{\varphi }^{\ast }\left( \nu \right) $ in the Hilbert space 
$L^{2}\left( \Omega \right) $ as antilinear map $\varphi \mapsto \varphi
^{\star },\varphi ^{\star \star }=\varphi $ satisfying%
\begin{equation*}
\left\langle \varphi ^{\star }|\psi \right\rangle =\int \varphi \left( -\nu
\right) \psi \left( \nu \right) \mathrm{d}\nu =\int \varphi \left( \nu
\right) \psi \left( -\nu \right) \mathrm{d}\nu =\left\langle \psi ^{\star
}|\varphi \right\rangle \;\;\;\forall \varphi ,\psi \in L^{2}\left( \Omega
\right) \text{.}
\end{equation*}

We shall assume without loss of generality that the Plansherel measure $%
\mathrm{d}\nu $ on the null subset $\mathrm{N}_{0}=\left\{ {\nu :{\kappa }%
(\nu )=0=\widetilde{\kappa }(\nu )}\right\} $ is zero, taking $\mathrm{N}%
_{0}=\emptyset $ such that its support $\Omega $ is identified with the
union $\mathrm{N}_{+}^{\perp }\cup \mathrm{N}_{-}^{\perp }$, where 
\begin{equation*}
\mathrm{N}_{+}^{\perp }=\left\{ {\nu :\ {\kappa }(\nu )>0}\right\} ,\mathrm{N%
}_{-}^{\perp }=\left\{ {\nu :\widetilde{\kappa }(\nu )>0}\right\} .
\end{equation*}%
are the complementary subsets of the null sets $\mathrm{N}_{+}$ and $\mathrm{%
N}_{-}$ for $\kappa \in L^{1}\left( \mathrm{N}_{+}^{\perp }\right) $ and $%
\widetilde{\kappa }\in L^{1}\left( \mathrm{N}_{-}^{\perp }\right) $. Let us
prove that the pairs $\left\{ \Omega \ni \nu \mapsto \check{x}_{j}(\nu
)\right\} \in L^{2}\left( \mathrm{N}_{+}^{\perp }\right) $, $\left\{ \Omega
\ni \nu \mapsto \hat{x}_{j}(\nu )\right\} \in L^{2}\left( \mathrm{N}%
_{-}^{\perp }\right) $ of complex amplitudes 
\begin{equation*}
\check{x}_{j}(\nu )={\kappa }{(\nu )}^{1/2}u_{j}(\nu )\equiv \langle \nu |%
\mathrm{x}_{j}\rangle ,\ \ \ \hat{x}_{j}(\nu )=\widetilde{\kappa }{(\nu )}%
^{1/2}u_{j}(\nu )\equiv \langle \nu |\widetilde{\mathrm{x}}_{j}\rangle
\end{equation*}
which are related by the isometric involution $\star $ as%
\begin{equation*}
\hat{x}_{j}={\kappa (-\nu )}^{1/2}u_{j}(\nu )={\kappa (-\nu )}%
^{1/2}u_{j}^{\ast }(-\nu )=\check{x}_{j}^{\star }
\end{equation*}%
where $u^{\ast }=\overline{u}$, give the spectral representation of the
quantum noise vectors $|\mathrm{x}_{j}\rangle $ and $|\widetilde{\mathrm{x}}%
_{j}\rangle $.

Indeed, by virtue of the completeness of $\left\{ u_{\cdot }(\nu )\mid \nu
\in \Omega \right\} $ it follows that%
\begin{equation*}
\check{x}{_{i}^{\dagger }}\check{x}{_{j}}:=\int_{{}}^{{}}{\check{x}}_{i}(\nu
)^{\ast }\check{x}_{j}(\nu )\mathrm{d}\nu =\int_{{}}^{{}}\kappa (\nu )%
\overline{u}_{i}(\nu )u_{j}(\nu )\mathrm{d}\nu ={K}_{ij}
\end{equation*}%
\begin{equation*}
{{\hat{x}}_{i}^{\dagger }{\hat{x}}_{j}}:=\int_{{}}^{{}}{{\hat{x}}_{i}}(\nu
)^{\ast }{\hat{x}_{j}}(\nu )\mathrm{d}\nu =\int_{{}}^{{}}\widetilde{\kappa }%
(\nu )\overline{u}_{i}(\nu )u_{j}(\nu )\mathrm{d}\nu ={\ \widetilde{K}}_{ij}
\end{equation*}%
\begin{equation*}
\check{x}{_{i}^{\dagger }\hat{x}_{j}}:=\int_{{}}^{{}}{\ \check{x}}_{i}(\nu
)^{\ast }{\hat{x}_{j}}(\nu )\mathrm{d}\nu =\int_{{}}^{{}}\gamma (\nu )%
\overline{u}_{i}(\nu )u_{j}(\nu )\mathrm{d}\nu ={G}_{ij}
\end{equation*}%
where we used the invariance of the measure on $\Omega $ with respect to the
flip $\nu \mapsto -\nu $.

Let us define the antilinear maps $|\mathrm{z}\rangle \mapsto {|\mathrm{z}%
\rangle }^{\sharp }$, $|\mathrm{\tilde{z}}\rangle \mapsto |\mathrm{\tilde{z}}%
{\rangle }^{\flat }$ of vector conjugation 
\begin{equation*}
{\check{z}}^{\sharp }=\lambda ^{-1/2}\check{z}^{\star },\ \ \ {\hat{z}}%
^{\flat }=\lambda ^{1/2}{\hat{z}}^{\star },
\end{equation*}
respectively on the dense domains $\mathcal{D}_{\sharp }\subset L^{2}(%
\mathrm{N}_{+}^{\perp })$ and $\mathcal{D}_{\flat }\subset L^{2}\left( 
\mathrm{N}_{-}^{\perp }\right) $ as the subspaces 
\begin{equation*}
\mathcal{D}_{\sharp }=\left\{ \check{z}{:{\lambda }^{1/2}\check{z}\in {L}%
^{2}(\Omega )}\right\} ,\mathcal{D}_{\flat }=\left\{ {\hat{z}:{\lambda }%
^{-1/2}\hat{z}\in {L}^{2}(\Omega )}\right\}
\end{equation*}%
of square integrable complex amplitudes $\check{z}\in L^{2}(\Omega )$ with
the support in $\mathrm{N}_{+}^{\perp \text{ }}$ and $\hat{z}\in L^{2}\left(
\Omega \right) $ with the support in $\mathrm{N}_{-}^{\bot }$ where $\lambda
^{1/2}=\sqrt{\widetilde{\kappa }/\kappa }$ and $\lambda ^{-1/2}=\sqrt{\kappa
/\widetilde{\kappa }}$ are well defined as positive operators of
multiplications respectively by $\left[ {\widetilde{\kappa }(\nu )/{\kappa }%
(\nu )}\right] ^{1/2}$ and $\left[ {\kappa (\nu )/\widetilde{\kappa }(\nu )}%
\right] ^{1/2}$ into the subspace $L^{2}\left( \Theta \right) $ the common
support 
\begin{equation*}
\Theta =\left\{ \nu :\kappa \left( \nu \right) \neq 0\neq \widetilde{\kappa }%
(\nu )\right\} =\mathrm{N}_{+}^{\bot }\cap \mathrm{N}_{-}^{\bot }.
\end{equation*}

Note that $\mathcal{D}_{\sharp }\subseteq L^{2}(\Omega )$ is generated by
the complex amplitudes $\check{z}=\sum \zeta _{j}{\check{x}}_{j}$ with $%
\check{z}^{\star }\in \mathcal{D}_{\flat }$ and the adjoint conjugation on
the subspace $\mathcal{D}_{\flat }\subseteq L^{2}(\Omega )$ is generated by $%
\hat{z}=\sum \zeta _{j}{\hat{x}}_{j}$ with $\hat{z}^{\star }\in \mathcal{D}%
_{\sharp }$ such that $\mathcal{D}_{\sharp }=\mathcal{D}_{\flat }^{\star }$.
Moreover, $\check{z}^{\sharp \star }=\check{z}^{\star \flat }$, $\hat{z}%
^{\flat \star }=\hat{z}^{\star \sharp }$ and $\hat{z}^{\flat }=\lambda 
\check{z}^{\sharp }$, $\check{z}^{\sharp }=\lambda ^{-1}\check{z}^{\flat }$
are densely defined involutions in the subspace $L^{2}\left( \Theta \right) $
and $\lambda ^{\star }=\lambda ^{-1}$ on $\Theta $.

In the case of nonzero temperature, when $\Theta =\Omega $ and $\mathcal{D}%
_{\sharp }$, $\mathcal{D}_{\flat }$ are dense in $L^{2}\left( \Omega \right) 
$, the spectral representations of quantum noise and the output process are
connected by the spectral linear filters

\begin{equation*}
\hat{x}(\nu )={\lambda (\nu )}^{1/2}\check{x}(\nu ),\ \ \ \check{x}(\nu )={%
\lambda (\nu )}^{-1/2}\hat{x}(\nu ),
\end{equation*}%
and $\check{x}_{j}^{\sharp }=\check{x}_{j}$, ${{\hat{x}}^{\flat }}_{j}=\hat{x%
}_{j}$ for the generating spectral sequences ${\ }\left( \check{x}%
_{j}\right) $ and $\left( \hat{x}_{j}\right) $:%
\begin{equation*}
\check{x}_{j}^{\sharp }(\nu )=\lambda {(\nu )}^{-1/2}\hat{x}_{j}(\nu )=%
\check{x}_{j}(\nu ),\ \ \ \nu \in \mathrm{N}_{-}^{\perp },
\end{equation*}%
\begin{equation*}
{{\hat{x}}^{\flat }}_{j}(\nu )=\lambda {(\nu )}^{1/2}{\check{x}_{j}}(\nu )=\ 
{\hat{x}}_{j}(\nu ),\ \ \ \nu \in \mathrm{N}_{+}^{\perp }.
\end{equation*}
The latter implies $|\mathrm{z}^{\dag }\rangle ={|\mathrm{z}\rangle }%
^{\sharp },$ $|\mathrm{\tilde{z}}^{\dag }\rangle =|\mathrm{\tilde{z}}{%
\rangle }^{\flat }$ for the complex-linear combinations $\mathrm{z}=\sum
\zeta _{i}\mathrm{x}_{i}$, $\widetilde{\mathrm{z}}=\sum \zeta _{i}\widetilde{%
\mathrm{x}}_{i}$ since 
\begin{eqnarray*}
\left\langle \mathrm{x}_{i}^{\dag }|\mathrm{x}_{j}^{\dag }\right\rangle
&=&\left\langle \mathrm{x}_{i}|\mathrm{x}_{j}\right\rangle =\left\langle 
\widetilde{\mathrm{x}}_{j}|\widetilde{\mathrm{x}}_{i}\right\rangle =\hat{x}%
_{j}^{\dagger }\hat{x}_{i}=\check{x}_{i}^{\dagger }\check{x}_{j}=\check{x}%
_{i}^{\sharp \dagger }\check{x}_{j}^{\sharp }, \\
\left\langle \widetilde{\mathrm{x}}_{i}^{\dag }|\widetilde{\mathrm{x}}%
_{j}^{\dag }\right\rangle &=&\left\langle \widetilde{\mathrm{x}}_{i}|%
\widetilde{\mathrm{x}}_{j}\right\rangle =\left\langle \mathrm{x}_{j}|\mathrm{%
x}_{i}\right\rangle =\check{x}_{j}^{\dagger }\check{x}_{i}=\hat{x}%
_{i}^{\dagger }\check{x}_{j}=\hat{x}_{i}^{\flat \dagger }\hat{x}_{j}^{\flat
}, \\
\left\langle \widetilde{\mathrm{x}}_{i}^{\dag }|\mathrm{x}_{j}^{\dag
}\right\rangle &=&\left\langle \widetilde{\mathrm{x}}_{i}|\mathrm{x}%
_{j}\right\rangle =\left\langle \mathrm{x}_{j}|\widetilde{\mathrm{x}}%
_{i}\right\rangle =\check{x}_{j}^{\dagger }\hat{x}_{i}=\hat{x}_{i}^{\dagger }%
\check{x}_{j}=\hat{x}_{i}^{\flat \dagger }\check{x}_{j}^{\sharp }.
\end{eqnarray*}

The spectral amplitudes $\check{x}_{j}$, $\hat{x}_{j}$ in the time
representation 
\begin{equation*}
{x_{j}}^{i}=\int_{{}}^{{}}{\overline{u}}_{i}(\nu )\check{x}_{j}(\nu )\mathrm{%
d}\nu \ ,\ \ \ \ {\ \widetilde{x}_{j}}^{\ i}=\int_{{}}^{{}}{\overline{u}}%
_{i}(\nu ){\hat{x}}_{j}(\nu )\mathrm{d}\nu
\end{equation*}%
describe the canonical vector realization $x_{j},{\widetilde{x}}_{j}$ of $|%
\mathrm{x}_{j}\rangle ,|{\widetilde{\mathrm{x}}}_{j}\rangle $. It is given
by the matrix elements of the square roots $X=K^{1/2}$ and $\widetilde{X}={%
\widetilde{K}}^{1/2}$ as

\begin{equation*}
{x_{j}}^{i}={X}_{ij},\ \ \ {{\widetilde{x}}_{j}}^{\ i}={\widetilde{X}}%
_{ij},\ \ \ i,j\in \mathbf{Z}
\end{equation*}%
such that ${{\widetilde{x}}_{j}}^{i}={{\overline{x}}_{j}}^{i}$ as $%
\widetilde{X}=\overline{X}$. They are thought as the ${\ell }^{2}(\mathbf{Z}%
) $ columns

\begin{equation*}
{x}_{j}={({x_{j}}^{i})}^{i\in \mathbf{Z}}\ ,\ \ \ \ {\widetilde{x}}_{j}={({{%
\overline{x}}_{j}}^{i})}^{i\in \mathbf{Z}}
\end{equation*}%
with the scalar products ${x}_{i}^{\dag }{x}_{j}=\langle \mathrm{x}_{i}|%
\mathrm{x}_{j}\rangle ,{\widetilde{x}}_{i}^{\dag }{\widetilde{x}}%
_{j}=\langle {\widetilde{\mathrm{x}}}_{i}|{\widetilde{\mathrm{x}}}%
_{j}\rangle $:

\begin{equation*}
{{x}_{i}}^{\dag }{x}_{j}=\sum\limits_{k\in \mathbf{Z}}^{{}}{\overline{X}}%
_{ki}{X}_{kj}=K_{ij}=\sum\limits_{k\in \mathbf{Z}}^{{}}{X}_{ki}{\overline{X}}%
_{kj}={{\ \widetilde{x}}_{j}}^{\dag }{\widetilde{x}}_{i},\ \ \ {{x}_{i}}%
^{\dag }{\widetilde{x}}_{j}=G_{ij}.
\end{equation*}%
Hence the inversion%
\index{time reversion} ${x}\mapsto {%
\widetilde{x}}$ is represented by the usual complex conjugation $\widetilde{x%
}=\overline{x},{\ \overline{x}_{j}}={({\ {\overline{x}}_{j}}^{i})}^{i\in 
\mathbf{Z}}$, which coincides with transposition : ${{\overline{x}}_{j}}^{i}=%
{x_{i}}^{j}$ due to the selfadjointness of the square roots of $K=K^{\dag }$
and so ${\overline{K}}={\ \overline{K}}^{\dag }$. Although the vectors ${x}%
_{j},{\widetilde{x}}_{j}$ seem to be complex if ${\widetilde{x}}_{j}\neq {x}%
_{j}$ due to ${x}_{j}^{\ast }:=\overline{x_{j}}=\widetilde{{x}}_{j}$, they
are selfadjoint ${{x}_{j}}^{\sharp }={x}_{j},\ {{\ \widetilde{x}}_{j}}%
^{\flat }={\widetilde{x}}_{j}$ in the case $\Theta =\Omega $ with respect to
the involution ${z}^{\sharp }=L^{-1/2}{z}^{\ast },\ {\widetilde{z}}^{\flat
}=L^{1/2}{\widetilde{z}}^{\ast }$, where $z^{\ast }=\overline{z}$:

\begin{equation*}
L^{1/2}{x}_j=\sum \limits_{}^{}{L^{1/2}}_k{x_j}^k={\widetilde{x}}_j,\ \ \ \
\ L^{-1/2}{\widetilde{x}}_j=\sum \limits_{}^{}{L^{-1/2}}_k{x_j}^k={x}_j
\end{equation*}
are the linear input-output and reversed filters in the time representation.
The operator $L={\overline{K}}K^{-1}$ with the matrix elements

\begin{equation*}
L_{ij}=\int_{{}}^{{}}{\overline{u}}_{i}(\nu )\lambda (\nu )u_{j}(\nu )%
\mathrm{d}\nu ,\ \ \ \ \ \lambda (\nu )=\widetilde{\kappa }(\nu )/{\kappa }%
(\nu )
\end{equation*}%
is characterized by the properties $L>0,{\overline{L}}=L^{-1}$ and is called
the modular operator in this invertible case. It can be described by the
adjoint involutions as $L|\mathrm{z}\rangle ={|}${$\mathrm{z}$}${\rangle }%
^{\sharp \flat }$, and its algebraic analogue is the main object of study in
Tomita-Takesaki theory. Note that if $K$ is not invertible, the canonical
realizations of $\mathrm{x}$ and $\widetilde{\mathrm{x}}$ are still
connected by complex conjugation as the isometric involution {$\mathrm{z}%
\mapsto $}${\mathrm{z}}^{\ast }$, although it may not be described by the
linear filtering%
\index{linear filtering} because the adjoint involutions $\sharp ,\flat $
may not be represented in $\mathcal{E}$. In contrast to the classical theory
the constructed canonical realization ${%
\widetilde{x}}_{j}$ of the reversed quantum process does not coincide with ${%
x}_{j}$ and is even orthogonal to the real quantum noise if $\mathrm{N}%
_{+}\cup \mathrm{N}_{-}=\Omega $.

In the case of a stationary time sequence $x_j =\sqrt {\varepsilon}x (t_j ),
\ t_j=\varepsilon j$ the spectrum $\Omega$ is in the interval $[-{\varepsilon%
}^{-1} / 2, {\varepsilon}^{-1}/2] \subset \mathbf{R}$, the flip $\nu \mapsto
-\nu$ is the usual reflection, and

\begin{equation*}
u_{-j}(\nu )={\varepsilon }^{1/2}\exp \{2\pi \mathrm{i}\nu t_{j}\}=\overline{%
u_{j}}\left( \nu \right) .
\end{equation*}%
The canonical vector realizations of the stationary sequences $(x_{j})$ and $%
({\ \widetilde{x}}_{j})$ is given in the time representation by the matrix
elements ${x_{j}}^{i}=\chi _{j-i},\ \ {{\widetilde{x}}_{j}}^{\ i}=\widetilde{%
\chi }_{j-i}$, where the complex sequences

\begin{eqnarray*}
\chi _{j} &=&{\varepsilon }^{1/2}\int_{\Omega }^{{}}{{\kappa }(\nu )}%
^{1/2}\exp \{-2\pi \mathrm{i}\nu t_{j}\}\mathrm{d}\nu =\int_{\Omega
}x_{j}\left( \nu \right) \mathrm{d}\nu \\
{\widetilde{\chi }}_{j} &=&{\varepsilon }^{1/2}\int_{\Omega }^{{}}{%
\widetilde{\kappa }(\nu )}^{1/2}\exp \{-2\pi \mathrm{i}\nu t_{j}\}\mathrm{d}%
\nu =\int_{\Omega }\widetilde{x}_{j}\left( \nu \right) \mathrm{d}\nu
\end{eqnarray*}%
are connected by the usual time reflection ${\widetilde{x}}({t}_{j})=x(-{t}%
_{j})$ with respect to $t=0$: $\widetilde{\chi }_{j}=\chi _{-j}$. If $K$ is
invertible, this time reflection is described by the stationary input-output
and reversed linear complex filters

\begin{equation*}
\chi _{-i}=\sum\limits_{j\in \mathbf{Z}}^{{}}{l^{1/2}}_{i-j}{\chi _{j}},\ \
\ \ \ \chi _{i}=\sum\limits_{j\in \mathbf{Z}}^{{}}{l^{-1/2}}_{i-j}{\chi }%
_{-j},
\end{equation*}%
where ${l^{1/2}}_{i-j}={L^{1/2}}_{ij},\ \ \ {l^{-1/2}}_{i-j}={L^{-1/2}}_{ij}$
are given by the integrals

\begin{equation*}
{{l_{\Theta }}^{1/2}}_{j}:=\varepsilon \int_{\Theta }^{{}}{\lambda (\nu )}%
^{1/2}\exp \{2\pi \mathrm{i}\nu {t}_{j}\}\mathrm{d}\nu ,\ \ \ {{l_{\Theta }}%
^{-1/2}}_{j}:=\varepsilon \int_{\Theta }^{{}}{\lambda (\nu )}^{-1/2}\exp
\{2\pi \mathrm{i}\nu {t}_{j}\}\mathrm{d}\nu
\end{equation*}%
over the support $\Theta \subseteq \Omega $ of $\gamma =\left( \kappa 
\widetilde{\kappa }\right) ^{1/2}\ $in the general case.


\section{Spectral Decomposition and Linear Filtering%
\index{linear filtering}}

Let us consider the stationary quantum noise $\mathrm{x}:t\mapsto \mathrm{x}%
(t)$ with continuous time $t\in \mathbf{R}$. It can be treated as the limit
as $\varepsilon \rightarrow 0$ of a stationary sequence $\varepsilon ^{-1/2}$%
{{$\mathrm{x}$}}${^{\varepsilon }}_{j}$ with correlations ${{k}^{\varepsilon
}}_{j}=\varepsilon k({t})$, given by a complex positive-definite function 
\begin{equation*}
k({t})=\int_{-\infty }^{\infty }\exp \left\{ {\ 2\pi \mathrm{i}\nu t}%
\right\} \kappa (\nu )\mathrm{d}\nu =\left\langle {\mathrm{x}(t)|\mathrm{x}%
(0)}\right\rangle .
\end{equation*}%
This may exist only as a generalized function(distribution) if the support $%
\mathrm{N}_{+}^{\perp }\subseteq \mathbf{R}$ of the spectral density $\kappa 
$ is unbounded. The reversed noise $%
\widetilde{\mathrm{x}}:t\mapsto \widetilde{\mathrm{x}}(t)$, corresponding to
the spectral density $\widetilde{\kappa }(\nu )=\kappa (-\nu )$, is
described up to second order by the autocorrelation function 
\begin{equation*}
\widetilde{k}({t})=\int_{-\infty }^{\infty }\exp \left\{ {2\pi \mathrm{i}\nu
t}\right\} \widetilde{\kappa }(\nu )\mathrm{d}\nu =\left\langle {\widetilde{%
\mathrm{x}}({t})|\widetilde{\mathrm{x}}(0)}\right\rangle
\end{equation*}%
and by the symmetric cross-correlation function 
\begin{equation*}
r(t)=\int_{-\infty }^{\infty }\exp \left\{ {\ 2\pi \mathrm{i}\nu t}\right\}
\gamma (\nu )\mathrm{d}\nu =\left\langle {{\mathrm{x}}({t})|\widetilde{%
\mathrm{x}}(0)}\right\rangle ,
\end{equation*}%
where $\gamma (\nu )={\left[ {\widetilde{\kappa }(\nu )\kappa (\nu )}\right] 
}^{1/2}$, so that $r$ is the convolutional square root of 
\begin{equation*}
\left[ {\widetilde{k}\ast k}\right] ({t})=\int_{-\infty }^{\infty }k(s-t)k(s)%
\mathrm{d}s.
\end{equation*}%
If $\widetilde{k}\ast k=0$, i.e. if $\widetilde{\kappa }\kappa =0$, and so $%
\mathrm{N}_{+}^{\perp }\cap \mathrm{N}_{-}^{\perp }=\emptyset $ we have 
\emph{stationary vacuum noise}. The stationary noise is called \emph{%
standard vacuum noise} if $k({t})+\widetilde{k}({t})$ is the Dirac $\delta $%
-function, i.e. if $\widetilde{\kappa }+\kappa =1$.

Such a noise is purely nonclassical because the condition $\widetilde{\kappa 
}\kappa =0$ for a symmetric function $\widetilde{\kappa }=\kappa $ is only
possible in the trivial case $\mathrm{N}_{+}^{\perp }=\emptyset =\mathrm{N}%
_{-}^{\perp }$ when $\mathrm{x}=0=\widetilde{\mathrm{x}}$.

In the general case the subset $\Theta =\mathrm{N}_{+}^{\perp }\cap \mathrm{N%
}_{-}^{\perp }$ of $\Omega $ is not empty, but the complement ${\Theta }%
^{\perp }=\Omega \setminus \Theta $ can be decomposed into the disjoint
union of 
\begin{equation*}
\mathrm{N}_{-}=\left\{ {\nu \in \Omega |\widetilde{\kappa }(\nu )=0}\right\}
\subseteq \mathrm{N}_{+}^{\perp },\ \ \mathrm{N}_{+}=\left\{ {\nu \in \Omega
|\kappa (\nu )=0}\right\} \subseteq \mathrm{N}_{-}^{\perp },\ \ \mathrm{N}%
_{+}\cap \mathrm{N}_{-}=\emptyset .
\end{equation*}%
This means that the quantum noise and its time-reversed version can be
uniquely decomposed in the correlation theory into sums $\mathrm{x}_{o}+%
\mathrm{x}_{\Theta },\widetilde{\mathrm{x}}_{o}+\widetilde{\mathrm{x}}%
_{\Theta }$ of uncorrelated vacuum $(\mathrm{x}_{o},\widetilde{\mathrm{x}}%
_{o})$ and thermal $(\mathrm{x}_{\Theta },{\widetilde{\mathrm{x}}}_{\Theta
}) $ components. In the spectral representation they are given by 
\begin{equation*}
\check{x}_{o}(t)=P_{-}^{\perp }\check{x}(t),\ \ \ {\hat{x}}%
_{o}(t)=P_{+}^{\perp }\hat{x}(t),
\end{equation*}%
\begin{equation*}
\check{x}_{\Theta }(t)=P_{\Theta }\check{x}(t),\ \ \ {\hat{x}}_{\Theta
}(t)=P_{\Theta }\hat{x}(t),\ \ \ \ \ \forall t\in \mathbf{R}
\end{equation*}%
where ${P}_{+}^{\perp }$ and ${P}_{-}^{\perp }$ are the projectors on $L^{2}(%
\mathrm{N}_{+})$ and $L^{2}(\mathrm{N}_{-})$ and ${P}_{\Theta }={P}_{+}{P}%
_{-}$ is the orthoprojector on the subspace ${L}^{2}(\Theta )$. The
orthoprojectors $P_{+}=1-1_{+}$ and $P_{-}=1-1_{-}$ define the best
input-output and output-input linear estimates 
\begin{equation*}
P_{+}\hat{x}(t)=1_{\Theta }\hat{x}(t)={\lambda _{\Theta }}^{1/2}\check{x}%
(t)\;\;\text{of\ }\;\hat{x}\left( t\right) ={\lambda _{\Theta }}^{1/2}\check{%
x}(t)+\hat{x}_{o}\left( t\right) ,
\end{equation*}%
\begin{equation*}
P_{-}\check{x}(t)=1_{\Theta }\check{x}(t)={\lambda _{\Theta }}^{-1/2}\hat{x}%
(t)\;\;\text{of}\;\;\check{x}(t)={\lambda _{\Theta }}^{-1/2}\hat{x}(t)+%
\check{x}_{o}\left( t\right)
\end{equation*}%
where $1_{\Theta }$, defined by ${1}_{\Theta }(\nu )=1$ if $\nu \in \Theta $%
, ${1}_{\Theta }(\nu )=0$ if $\nu \not\in \Theta $, is the characteristic
function of $\Theta =\mathrm{N}_{+}^{\perp }\cap \mathrm{N}_{-}^{\perp }$, ${%
1}_{+}$ and ${1}_{-}$ are the characteristic functions of $\mathrm{N}_{+}$
and $\mathrm{N}_{-}$ , and 
\begin{equation*}
{\lambda }_{\Theta }(\nu )^{-1/2}=\left( {\sigma (\nu )/\widetilde{\sigma }%
(\nu )}\right) 1_{\Theta }(\nu )
\end{equation*}%
\begin{equation*}
{\lambda }_{\Theta }(\nu )^{1/2}=\left( {\widetilde{\sigma }(\nu )/\sigma
(\nu )}\right) 1_{\Theta }(\nu ).
\end{equation*}%
In the time representation 
\begin{equation*}
x(t)=\left\{ {\chi (t-s)|s\in \mathbf{R}}\right\} ,\ \ \ \widetilde{x}%
(t)=\left\{ {\widetilde{\chi }(t-s)|s\in \mathbf{R}}\right\}
\end{equation*}%
\begin{equation*}
\chi (t-s)=\int_{\Omega }^{{}}{\kappa (\nu )}^{1/2}\exp \left\{ {2\pi 
\mathrm{i}\nu (s-t)}\right\} \mathrm{d}\nu =\widetilde{\chi }(s-t)
\end{equation*}%
$x_{\Theta }(t)$ and ${\widetilde{x}}_{\Theta }(t)$ are obtained by the
replacing the interval $\Omega $ of integration in $\chi =\chi _{\Omega }$
and $\widetilde{\chi }=\widetilde{\chi }_{\Omega }$ by $\Theta $, with $%
x_{o}(t)$ given by $\chi _{{\mathrm{N}_{-}}}$ and ${\widetilde{x}}_{o}(t)$
given by $\chi _{{\mathrm{N}_{+}}}$. The optimal input-output and
output-input filters are given in the Fourier representation by the complex
stationary linear nonadapted integrals 
\begin{eqnarray*}
\left[ P_{+}\widetilde{x}(t)\right] (s) &=&\int_{-\infty }^{\infty }{%
l_{\Theta }}^{1/2}(s-r)\chi (t-r)\mathrm{d}r\equiv \left( l_{\Theta
}^{1/2}\ast x\right) \left( s\right) \ \  \\
\left[ P_{-}x(t)\right] (s) &=&\int_{-\infty }^{\infty }{l_{\Theta }}%
^{-1/2}(s-r)\widetilde{\chi }(t-r)\mathrm{d}r\equiv \left( l_{\Theta
}^{-1/2}\ast \widetilde{x}\right) \left( s\right) .
\end{eqnarray*}%
Here ${l_{\Theta }}^{1/2}(s-t)={L_{\Theta }}^{1/2}(s,t),\ {l_{\Theta }}%
^{-1/2}(s-t)={L_{\Theta }}^{-1/2}(s,t)$ are the complex positive definite
generalized functions 
\begin{equation*}
{l_{\Theta }}^{1/2}(t)=\int \lambda _{\Theta }{(\nu )}^{1/2}\exp \left\{ {%
2\pi \mathrm{i}\nu t}\right\} \mathrm{d}\nu ,\ \ \ {l_{\Theta }}%
^{-1/2}(t)=\int \lambda _{\Theta }{(\nu )}^{-1/2}\exp \left\{ {2\pi \mathrm{i%
}\nu t}\right\} \mathrm{d}\nu
\end{equation*}%
characterized by the modular property 
\begin{equation*}
{l_{\Theta }}^{1/2}(-t)={\overline{l}_{\Theta }}^{1/2}(t)={l_{\Theta }}%
^{-1/2}(t).
\end{equation*}%
The case $\mathrm{N}_{+}=\emptyset =\mathrm{N}_{-}$ corresponds to the
purely thermal noise which is called \emph{standard thermal noise}, if $%
\widetilde{\kappa }\kappa =1$ in a given spectral region $\Omega \subseteq 
\mathbf{R}$. If $\Omega =\mathbf{R}$, it can be written as $\left[ {\ {%
\widetilde{k}}\ast k}\right] (t)=\delta (t)$ in terms of the Dirac
correlation function $r(t)=\delta (t)$.

The thermal noise is called \emph{white} if the spectrum $\kappa $ and thus
also $\widetilde{\kappa }$ is flat: $\kappa (\nu )={\sigma }^{2}=\widetilde{%
\kappa }(\nu ),\ \forall \nu \in \Omega $. White noise coincides with its
time-reversed version and is essentially classical (at least in second
order). There exists just one such standard noise $\mathrm{x}=\mathrm{w}=%
\widetilde{\mathrm{x}}$ described in the second order theory by the
correlation function $k(t)=\delta (t)=\overline{k}(t)$.

In the quantum case there are many standard thermal noises and $\delta $%
-correlated with them reversed version $\widetilde{\chi }(t)=\chi (-t)$.
They are parametrized by a modular spectral function 
\begin{equation*}
\lambda (\nu )>0,\ \widetilde{\lambda }(\nu )=\lambda (-\nu )=\lambda (\nu
)^{-1},\ \forall \nu \in \Omega
\end{equation*}%
giving the standard correlation functions $k$, $\overline{k}$ as the
convolution square roots 
\begin{equation*}
k(t)={l}^{-1/2}(t),\ \overline{k}(t)={l}^{1/2}(t),\ \forall t\in \mathbf{R}
\end{equation*}%
Any stationary quantum process $\mathrm{y}:t\rightarrow \mathrm{y}(t)$ and
its reverse $\widetilde{\mathrm{y}}:t\rightarrow \widetilde{\mathrm{y}}(t)$
described up to second order by the correlation function 
\begin{equation*}
\langle \mathrm{y}(t)\mathrm{y}(0)\rangle =\int_{{}}^{{}}\exp \{2\pi \mathrm{%
i}\nu t\}{\sigma (\nu )}^{2}\mathrm{d}\nu =\langle \widetilde{\mathrm{y}}(0)%
\widetilde{\mathrm{y}}(t)\rangle
\end{equation*}%
can be obtained by stationary filtering of standard quantum noises $\mathrm{x%
}$ and $\widetilde{\mathrm{x}}$, i.e. such that their spectral densities 
\begin{equation*}
\kappa (\nu )={1}_{-}(\nu )+{{\lambda }_{\Theta }(\nu )}^{-1/2},\widetilde{%
\kappa }(\nu )={1}_{+}(\nu )+{{\lambda }_{\Theta }(\nu )}^{1/2}
\end{equation*}%
and $\gamma (\nu )={1}_{\Theta }(\nu )$. In the spectral representation 
\begin{equation*}
\check{y}(t)=\left\{ {\check{y}(t,\nu )|\nu \in \Omega }\right\} ,\ \ \ \hat{%
y}(t)=\left\{ {\hat{y}(t,\nu )|\nu \in \Omega }\right\}
\end{equation*}%
the filtering is given by 
\begin{equation*}
\check{y}(t,\nu )=f(\nu )\check{x}(t,\nu ),\ \hat{y}(t,\nu )=f(\nu )\hat{x}%
(t,\nu )
\end{equation*}%
where $\overline{f}=f=\widetilde{f}$ is a real symmetric transmission
function $f:\Omega \rightarrow \mathbf{R}_{+}$ of the standard complex
amplitudes 
\begin{equation*}
\check{x}(t,\nu )=\exp \{{2\pi \mathrm{i}\nu t}\}\check{x}(\nu ),\ \hat{x}%
(t,\nu )=\exp \{{2\pi \mathrm{i}\nu t}\}\hat{x}(\nu )
\end{equation*}%
\begin{equation*}
\check{x}(\nu )={1}_{-}(\nu )+{{\lambda }_{\Theta }(\nu )}^{-1/4},\ {\hat{x}}%
(\nu )={1}_{+}(\nu )+{{\lambda }_{\Theta }(\nu )}^{1/4}.
\end{equation*}%
The \emph{transmission function} $f$ is uniquely defined in the space ${L}%
^{2}(\Omega )$ by 
\begin{equation*}
f(\nu )={[\sigma (\nu )\widetilde{\sigma }(\nu )]}^{1/2}{1}_{\Theta }(\nu )+{%
[\sigma (\nu )\vee \widetilde{\sigma }(\nu )]}^{1/2}{1}_{\Theta }^{\perp
}(\nu ).
\end{equation*}%
In the time representation $y(t)=\{\psi (t-s)\},\ \widetilde{y}(t)=\{%
\widetilde{\psi }(t-s)\}$, given by the Fourier integral 
\begin{equation*}
\psi (t)=\int_{{}}^{{}}\hat{y}(\nu )\exp \{{2\pi \mathrm{i}\nu t}\}\mathrm{d}%
\nu ,\;\widetilde{\psi }(t)=\int_{{}}^{{}}\check{y}(\nu )\exp \{{2\pi 
\mathrm{i}\nu t}\}\mathrm{d}\nu ,
\end{equation*}%
the filters are written as real nonadapted stationary transformations 
\begin{equation*}
\psi (t-s)=\int_{-\infty }^{\infty }\phi (s-r)\chi (t-r)\mathrm{d}r=%
\widetilde{\psi }(s-t)
\end{equation*}%
where 
\begin{equation*}
\phi (t)=\int_{{}}^{{}}f(\nu )\exp \{{2\pi \mathrm{i}\nu t}\}\mathrm{d}\nu .
\end{equation*}

An important example of the stationary quantum thermal noise is given by the
Planck's spectral density 
\begin{equation*}
\kappa (\nu )=h\nu {(\mathrm{e}^{\beta h\nu }-1)}^{-1},\ \widetilde{\kappa }%
(\nu )=h\nu {(1-\mathrm{e}^{-\beta h\nu })}^{-1}.
\end{equation*}%
In this case $\widetilde{\kappa }(\nu )-\kappa (\nu )=h\nu >0$ if $\nu >0$,
i.e. $\Theta _{+}=\mathbf{R}_{+}$, $\lambda (\nu )=\exp \{\beta h\nu \}\neq
0\ \forall \nu \in \mathbf{{R},}$ and $\lambda (\nu )\neq 1$ for $\nu \neq 0$%
, and $\lambda (\nu )=1$ as for the classical white noise only if $\beta =0$%
. It describes stationary Bose noise in which the electromagnetic (optical)
field is in an equilibrium state of the temperature $\theta =1/\beta $. The
modular function for such noise, 
\begin{equation*}
l(t)=\int_{{}}^{{}}\exp \{{2\pi \mathrm{i}\nu t}\}\exp \{\beta h\nu \}%
\mathrm{d}\nu =\delta \left( t+\frac{\hbar }{\mathrm{i}}\beta \right) =\exp
\left\{ \frac{\hbar }{\mathrm{i}}\beta {\frac{\mathrm{d}}{{\mathrm{d}t}}}%
\right\} \delta (t),
\end{equation*}%
is extremely singular, as is the correlation function 
\begin{equation*}
k(t)=\frac{\hbar }{\mathrm{i}}{\left( {\exp \left\{ \frac{\hbar }{\mathrm{i}}%
\beta {\ \frac{\mathrm{d}}{{\mathrm{d}t}}}\right\} -1}\right) }^{-1}{\frac{%
\mathrm{d}}{{\mathrm{d}t}}}\ \delta (t)
\end{equation*}%
($\hbar =h/2\pi $) with complex conjugate

\begin{equation*}
\overline{k}(t)=\frac \hbar {\mathrm{i}}{\left( 1-\exp \left\{ -\frac \hbar {%
\mathrm{i}}\beta {\frac{\mathrm{d}}{{\mathrm{d}t}}}\right\} \right) }^{-1}{\ 
\frac{\mathrm{d}}{{\mathrm{d}t}}}\ \delta (t)
\end{equation*}
for the time-reversed Planck noise corresponding to the density $\widetilde{%
\kappa }$.

Although the cross correlation function

\begin{equation*}
r(t)=\frac{\hbar }{2}{\left[ \sin \left( \frac{\hbar }{2}\beta {\ \frac{%
\mathrm{d}}{{\mathrm{d}t}}}\right) \right] }^{-1}{\frac{\mathrm{d}}{{\mathrm{%
d}t}}}\delta (t)
\end{equation*}%
is less singular, due to the integrability of its spectral density

\begin{equation*}
\gamma (\nu )=h\nu {[\exp \{\beta h\nu /2\}-\exp \{-\beta h\nu /2\}]}^{-1},
\end{equation*}%
quantum stochastic integrals corresponding to linear stationary filters for
such noise are not defined in the usual sense. Indeed, the standard quantum
pair $(\mathrm{x},\widetilde{\mathrm{x}})$ corresponding to the Planck
density have the singular correlation functions

\begin{equation*}
{l}^{-1/2}(t)=\delta (t+\mathrm{i}\hbar \beta /2),\ \ \ \ \ {l}%
^{1/2}(t)=\delta (t-\mathrm{i}\hbar \beta /2)
\end{equation*}%
and the cross-correlation function $r(t)=\delta (t)$. To obtain the Planck
noise $\mathrm{y}$ and its time-reversed version $\widetilde{\mathrm{y}}$
one should integrate the standard quantum noise $\mathrm{x}$ and $\widetilde{%
\mathrm{x}}$ with the non singular filtering function

\begin{equation*}
f(t)=\int_{}^{}{\left[ h\nu /2 \mathrm{sh}(\beta h\nu /2)\right] }^{1/2}\exp
\{{2\pi \mathrm{i}\nu t}\}\mathrm{d}\nu
\end{equation*}
But the result of the integration remains singular, corresponding to the
singularity of $k$ and $\overline{k}$.

\section{Quantum Stochastic Integration}

In quantum theory the noise $\mathrm{x}\left( t\right) $ is thought of as
represented by Hermitian, or even selfadjoint operators \ in a Hilbert space
of state vectors (probability amplitudes), with zero mean values $\phi \left[
\mathrm{x}\left( t\right) \right] =\left\langle {\phi |\mathrm{x}(t)|\phi }%
\right\rangle =0$. Here $|\phi \rangle $ is a fixed unit vector of the given
state $\phi $ which can be always be chosen as vector state represented by
the vacuum vector$\ |\left. {\phi }\right\rangle =\delta _{\emptyset }$ in a
Fock space $\mathcal{H}=\mathcal{F}$. The quantum correlations%
\index{quantum correlation} $K(s,t)=k(s-t)$ are defined for a stationary
state $\phi $ by the scalar product in $\mathcal{H}$,%
\begin{equation*}
k(t)=\left\langle {\phi }\right. |\mathrm{x}(t)\mathrm{x}(0)|\left. {\phi }%
\right\rangle \equiv \phi \left[ \mathrm{x}\left( t\right) \mathrm{x}\left(
0\right) \right] .
\end{equation*}%
If $k(t)$ is complex, the operator $\mathrm{x}\left( t\right) =${$\mathrm{x}$%
}${(t)}^{\dag }$ will not commute with $\mathrm{x}(0)=${$\mathrm{x}$}${(0)}%
^{\dag }$. The time-reversed correlations $%
\widetilde{k}(t)=k(-t)$ are \textquotedblleft observable\textquotedblright\
in the real arrow of time only for the time-reversed noise $\widetilde{%
\mathrm{x}}$ which gives a natural representation in the same Hilbert space $%
\mathcal{H}$ an output process $\widetilde{\mathrm{x}}:t\mapsto \widetilde{%
\mathrm{x}}(t)$, for which $\widetilde{k}(t)=\left\langle {\phi }\right. |%
\widetilde{\mathrm{x}}(t)\widetilde{\mathrm{x}}(0)|\left. {\phi }%
\right\rangle $ with the same $\phi \in \mathcal{H}$.

The vector processes $t\mapsto |\mathrm{x}(t)\rangle $, $t\mapsto |%
\widetilde{\mathrm{x}}{(}t{)\rangle }$ are obtained from the operator
representations in $\mathcal{H}$ by the identifications 
\begin{equation*}
|\left. \mathrm{x}{(t)}\right\rangle :=\mathrm{x}(t)|\left. {\phi }%
\right\rangle \equiv x\left( t\right) ,\ \ \ |\widetilde{\mathrm{x}}\left. {%
(t)}\right\rangle :=\widetilde{\mathrm{x}}(t)|\left. {\phi }\right\rangle
\equiv \widetilde{x}\left( t\right)
\end{equation*}%
in the minimal subspace generated by the set $\left\{ {\mathrm{x}(t)|\left. {%
\phi }\right\rangle ,\widetilde{\mathrm{x}}(t)|\left. {\phi }\right\rangle }%
\right\} \subset \mathcal{H}$. In a classical theory this gives a one-to-one
correspondence between vectors and operators of multiplication in the total
space $\mathcal{H}$. This is not the case in quantum probability theory as
one can observe for vacuum correlations, for which it is possible that $|${$%
\mathrm{z}$}${\rangle }=\mathrm{z}|{\phi \rangle }=0$ for annihilation
operators $\mathrm{z}\neq 0$. This is why the vector integration theory is
not sufficient in the quantum case; we need an operator form of quantum
stochastic integration.

According to the nondemolition (causality) principle of quantum theory \cite%
{bib:noise11} the input observables $\mathrm{x}(0)$ must commute with the
fundamental output observables $\widetilde{\mathrm{x}}(t)$ for all $t\leq 0$
and they must commute also for $t\geq 0$ because the noise $\mathrm{x}$ can
be regarded as the output process for the time-reversed noise $\widetilde{%
\mathrm{x}}$. At the correlation level this is expressed in terms of the
reflection symmetry 
\begin{equation*}
\left\langle {\phi }\right. |\mathrm{x}(0)\widetilde{\mathrm{x}}(t)|\left. {%
\phi }\right\rangle =r(t)=\left\langle {\phi }\right. |\widetilde{\mathrm{x}}%
(t)\mathrm{x}(0)|\left. {\phi }\right\rangle ,\ \ \ \forall t\in \mathbf{R}
\end{equation*}%
for the cross correlation function $r(t)={(\widetilde{k}\ast k)}^{1/2}(t)$.

Now we are going to develop a second order theory of quantum stochastic
integration with respect to any colored quantum noise and its fundamental
output process. The realization of such theory was given in \cite{bib:noise8}
in terms of a representation of the canonical commutation relations in the
symmetric Fock space $\mathcal{H}$. This provides a one-to-one
correspondence between second order integration theory and the Gaussian
integration theory of quantum thermal noise with respect to the vacuum state 
$\phi $. Without loss of generality we may assume that the pare $\mathrm{x}%
\left( t\right) ,\widetilde{\mathrm{x}}\left( t\right) $ is standard, i.e. $%
\kappa \widetilde{\kappa }=1$.

Let us denote by $\zeta $ the complex test functions $t\mapsto \zeta (t)$,
for which the \textquotedblleft integrals\textquotedblright\ 
\begin{equation*}
\zeta \cdot \mathrm{x}=\int_{-\infty }^{\infty }\zeta (t)\mathrm{x}(t)%
\mathrm{d}t,\ \ \ \zeta \cdot \widetilde{\mathrm{x}}=\int_{-\infty }^{\infty
}\zeta (t)\widetilde{\mathrm{x}}(t)\mathrm{d}t
\end{equation*}%
are expected to be well defined as selfadjoint operators in $\mathcal{H}$ if 
$\overline{\zeta }=\zeta $. If $\zeta $ is given as a Fourier integral 
\begin{equation*}
\check{c}(t):=\int_{{}}^{{}}c(\nu )\exp \left\{ {2\pi \mathrm{i}\nu t}%
\right\} \mathrm{d}\nu \equiv \hat{c}(-t),
\end{equation*}%
then the operator-valued functionals $\zeta \mapsto \zeta \cdot \mathrm{x}$, 
$\zeta \mapsto \zeta \cdot \widetilde{\mathrm{x}}$ can be treated as quantum
stochastic integrals 
\begin{equation*}
\check{c}\cdot \mathrm{x}=\int_{{}}^{{}}c(\nu )\mathrm{\check{X}}(\mathrm{d}%
\nu ),\ \ \ \int_{{}}^{{}}c(\nu )\mathrm{\hat{X}}(\mathrm{d}\nu )=\hat{c}%
\cdot \mathrm{x}
\end{equation*}%
for the Fourier transforms $c=\widehat{\zeta }$ of $\zeta =\check{c}$ and $c=%
\check{\zeta}$ of $\zeta =\hat{c}$.

The quantum stochastic integrators $\mathrm{\check{X}}$, $\mathrm{\hat{X}}$,
defined in \cite{bib:noise13} as operator-valued measures on the spectral
intervals $\Delta \subset \Omega $, in second order are described by
spectral vector-measures%
\begin{equation*}
|\mathrm{\check{X}}{(\Delta )\rangle :=\mathrm{\check{X}}{(\Delta )|\phi
\rangle }\equiv }\check{X}\left( \Delta \right) ,\;\;|\mathrm{\hat{X}}{%
(\Delta )\rangle :=\mathrm{\hat{X}}{(\Delta )|\phi \rangle }\equiv }\hat{X}%
\left( \Delta \right)
\end{equation*}%
which are related by isometric involution $\star =-\ast $ on the vector
space as $\hat{X}^{\star }=\check{X}$, i.e. $\hat{X}^{\ast }\left( -\Delta
\right) =\check{X}\left( \Delta \right) $.

They, together with $\check{X}\left( \Delta \right) ^{\dagger }=\langle {%
\mathrm{\check{X}}{(\Delta )|}}$, $\hat{X}\left( \Delta \right) ^{\dagger
}=\langle \mathrm{\hat{X}}\left( \Delta \right) |$ are assumed to satisfy
the following properties:\newline
(i) orthogonal $\sigma $-additivity for disjoint unions $\sum {\Delta }%
_{i}=\Delta $%
\begin{equation*}
\check{X}\left( \Delta \right) =\oplus \check{X}\left( \Delta _{i}\right)
,\;\;\;\hat{X}\left( \Delta \right) =\oplus \hat{X}\left( \Delta _{i}\right)
.
\end{equation*}%
(ii) absolute continuity: 
\begin{eqnarray}
{\normalsize \check{X}}{({\Delta })}^{\dagger }{{\normalsize \check{X}}%
(\Delta }^{\prime }{)} &=&\int_{\Delta \cap {\Delta }^{^{\prime
}}}^{{}}\kappa (\nu )\mathrm{d}\nu  \notag \\
\hat{X}{({\Delta })}^{\dagger }\hat{X}{(\Delta }^{\prime }{)}
&=&\int_{\Delta \cap {\Delta }^{^{\prime }}}^{{}}\widetilde{\kappa }(\nu )%
\mathrm{d}\nu  \notag \\
\check{X}{({\Delta })}^{\dagger }\hat{X}{(\Delta }^{\prime }{)}
&=&\int_{\Delta \cap {\Delta }^{^{\prime }}}^{{}}\gamma (\nu )\mathrm{d}\nu 
\notag
\end{eqnarray}%
(iii) selfadjointness on $\Theta =\mathrm{N}_{+}^{\bot }\cap \mathrm{N}%
_{-}^{\bot }$: 
\begin{equation*}
\check{X}{(\Delta )}^{\sharp }=\lambda ^{-1}\cdot \hat{X}{(\Delta )}=\check{X%
}(\Delta ),\ \ \hat{X}{(\Delta )}^{\flat }=\lambda \cdot \hat{X}{(\Delta )}=%
{\normalsize \hat{X}}(\Delta )\;\;\;\Delta \subseteq \Theta ,
\end{equation*}%
where $\gamma =\left( {{\widetilde{\kappa }\kappa }}\right) ^{1/2}$, $%
\lambda \left( \nu \right) =\left( \tilde{\kappa}/\kappa \right) ^{1/2}$ and 
$\lambda \cdot \hat{X}\left( \Delta \right) =\int_{\Delta }^{\oplus }\lambda
\left( \nu \right) \hat{X}\left( \mathrm{d}\nu \right) $.\ The $\sigma $%
-additivity makes it possible to define this integral as integral as an
orthogonal vector measure $\check{X}$ absolutely continuous with $\hat{X}$
on $\Theta $. The condition (ii) can be symbolically written in the form of
a multiplication table 
\begin{equation*}
\check{X}(\mathrm{d}{\nu })^{\dagger }\check{X}{(\mathrm{d}\nu }^{\prime }{)}%
={\delta }_{\nu {\nu }^{^{\prime }}}\kappa (\nu )\mathrm{d}\nu ,\ \ \ \check{%
X}(\mathrm{d}{\nu })^{\dagger }\hat{X}{(\mathrm{d}\nu }^{\prime }{)}%
^{\dagger }={\delta }_{\nu {\nu }^{^{\prime }}}\gamma (\nu )\mathrm{d}\nu
\end{equation*}%
\begin{equation*}
\hat{X}(\mathrm{d}{\nu })^{\dagger }\check{X}{(\mathrm{d}\nu }^{\prime }{)}={%
\delta }_{\nu {\nu }^{^{\prime }}}\gamma (\nu )\mathrm{d}\nu ,\ \ \ \hat{X}(%
\mathrm{d}{\nu })^{\dagger }\hat{X}{(\mathrm{d}\nu }^{\prime }{)}={\delta }%
_{\nu {\nu }^{^{\prime }}}\widetilde{\kappa }(\nu )\mathrm{d}\nu
\end{equation*}%
where ${\delta }_{\nu {\nu }^{^{\prime }}}=0$ if $\nu \neq {\nu }^{^{\prime
}},{\delta }_{\nu {\nu }^{^{\prime }}}={1}$ if $\nu ={\nu }^{^{\prime }}$,
and $\hat{X}\left( \mathrm{d}\nu \right) =\lambda \left( \nu \right) \check{X%
}\left( \mathrm{d}\nu \right) $. An operator realization of this table can
be given in the Fock space $\mathcal{F}$ with respect to the vacuum state
vector $|\left. {\phi }\right\rangle =\mathcal{H}$ \cite{bib:noise8}.

Using this multiplication table we obtain the isometry property 
\begin{equation*}
\left\langle {\phi }\right. |\mathrm{y}^{\dag }\mathrm{y}|\left. {\phi }%
\right\rangle =\int_{{}}^{{}}{\left\vert a{(\nu ){\kappa (\nu )}^{1/2}+c(\nu
)}\widetilde{{\kappa }}{{(\nu )}^{1/2}}\right\vert }^{2}\mathrm{d}\nu
=b^{\dagger }{b}
\end{equation*}%
of the map $\mathrm{y}\mapsto b=a{\kappa }^{1/2}+c{\widetilde{\kappa }}%
^{1/2} $ for the operators $\mathrm{y}=\check{a}\cdot \mathrm{x}+\check{c}%
\cdot \widetilde{\mathrm{x}}=\eta \cdot \mathrm{x}$ with $\eta =\check{a}+%
\hat{c}$ on the state vector $\phi $ into the spectral representation $b(\nu
)=\left\langle {\nu }\right. |\left. \mathrm{y}\right\rangle $ of $|\left. 
\mathrm{y}\right\rangle =|\left. \eta \cdot \mathrm{x}\right\rangle $. This
extends the quantum stochastic integral $\eta \cdot \mathrm{x}$ with $%
b^{\dagger }b=\int \left\vert b\left( \nu \right) \right\vert ^{2}\mathrm{d}%
\nu <\infty $ from the functions $\eta \left( t\right) $ with simple $\left(
a,c\right) $ to any functions $\eta =\check{a}+\hat{c}$ for which the
complex amplitude $b$ remains square integrable. To ensure also $%
\left\langle {\phi }\right. |\mathrm{yy}^{\dagger }|\left. {\phi }%
\right\rangle <\infty $, we must add the condition of square integrability $%
b^{\star }\in L^{2}(\Omega )$ for the amplitude $b^{\star }={a}^{\star }{%
\kappa }^{1/2}+{c}^{\star }{\widetilde{\kappa }}^{1/2}$, that is, we define
the quantum stochastic integral $\mathrm{y}$ together with $\mathrm{y}^{\dag
}$ only for the test functions $\check{a},\hat{c}\subset \mathcal{E}$. In
the classical case, $\widetilde{\kappa }=\kappa $ when $b^{\star }=b^{\ast }$%
, the square integrability of $b$ and $b^{\ast }$ are equivalent to that of $%
a$ and $c$ with respect to the symmetric measure $\widetilde{\kappa }\ 
\mathrm{d}\nu =\kappa \ \mathrm{d}\nu $. In the purely quantum case of
vacuum noise it is not so, and the conditions $b,{b}^{\star }\in
L^{2}(\Omega )$ are equivalent to the square integrability of $a$ and $c$
with respect to $\kappa \vee \widetilde{\kappa }\ \mathrm{d}\nu $.

In spectral representation the standard vacuum noise $\mathrm{x}\left(
t\right) $ together with the reversed process $\widetilde{\mathrm{x}}\left(
t\right) $ is described by the \emph{canonical} operator-valued measures 
\begin{equation*}
\mathrm{A}^{+}\left( \Delta \right) =\left\{ 
\begin{array}{ll}
\mathrm{\hat{X}}(\Delta ),\Delta \subset \mathrm{N}_{+}^{\perp } &  \\ 
\mathrm{\check{X}}(\Delta ),\Delta \subset \mathrm{N}_{-}^{\perp } & 
\end{array}%
\right. ,\ \ \ \ \mathrm{A}_{-}\left( \Delta \right) =\left\{ 
\begin{array}{ll}
\mathrm{\check{X}}(\Delta ),\Delta \subset \mathrm{N}_{+}^{\perp } &  \\ 
\mathrm{\hat{X}}(\Delta ),\Delta \subset \mathrm{N}_{-}^{\perp } & 
\end{array}%
\right.
\end{equation*}%
of independent creation $\mathrm{A}_{\Delta }^{+}=\mathrm{A}^{+}\left(
\Delta \right) $ and annihilation $\mathrm{A}_{-}^{\Delta }=\mathrm{A}%
_{-}\left( \Delta \right) $ such that 
\begin{equation*}
\left\langle {\phi }\right. |\mathrm{A}_{\Delta }^{+}=0,\ \;\mathrm{A}%
_{-}^{\Delta }|\left. {\phi }\right\rangle =0.
\end{equation*}%
It defines the standard operator vacuum measures $\mathrm{\hat{X}}$ and $%
\mathrm{\check{X}}$ as%
\begin{equation*}
\mathrm{\hat{X}}(\Delta )=\mathrm{A}_{-}\left( \Delta \cap \mathrm{N}%
_{-}^{\perp }\right) +\mathrm{A}^{+}\left( \Delta \cap \mathrm{N}_{+}^{\perp
}\right) ,\;\;\mathrm{\check{X}}(\Delta )=\mathrm{A}_{-}\left( \Delta \cap 
\mathrm{N}_{+}^{\perp }\right) +\mathrm{A}^{+}\left( \Delta \cap \mathrm{N}%
_{-}^{\perp }\right)
\end{equation*}%
on the intervals $\Delta \subset \Omega $ for the given state $|\left. {\phi 
}\right\rangle \in \mathcal{H}$. The canonical pair $(\mathrm{A}_{-}^{0},%
\mathrm{A}_{0}^{+})$ of the creation and annihilation measures is
characterized by the following properties of flip-adjointness 
\begin{equation*}
\mathrm{A}_{-}^{\Delta }{}^{\dag }=\mathrm{A}_{-\Delta }^{+}\equiv \mathrm{A}%
_{+}^{\Delta }{,\;\;}\mathrm{A}_{\Delta }^{+\dag }=\mathrm{A}_{-}^{-\Delta
}\equiv \mathrm{A}_{\Delta }^{-},
\end{equation*}%
and orthogonality of all products apart of $\mathrm{A}_{-}({\Delta })\mathrm{%
A}_{+}(\Delta ^{\prime })$ for $\Delta \cap {\Delta }^{^{\prime }}\neq
\emptyset $: 
\begin{equation*}
\left\langle {\phi }\right. |\mathrm{A}_{-}(-{\Delta })\mathrm{A}^{+}(\Delta
^{\prime })|\left. {\phi }\right\rangle =\nu (\Delta \cap {\Delta }%
^{^{\prime }}),\ \ \ \left\langle {\phi }\right. |\mathrm{A}^{+}(-{\Delta })%
\mathrm{A}_{-}(\Delta ^{\prime })|\left. {\phi }\right\rangle =0,
\end{equation*}%
\begin{equation*}
\left\langle {\phi }\right. |\mathrm{A}^{+}(-{\Delta })\mathrm{A}^{+}(\Delta
^{\prime })|\left. {\phi }\right\rangle =0,\ \ \ \left\langle {\phi }\right.
|\mathrm{A}_{-}(-{\Delta })\mathrm{A}_{-}(\Delta ^{\prime })|\left. {\phi }%
\right\rangle =0.
\end{equation*}%
This table can be written also symbolically in the canonical form \cite%
{bib:noise5} as 
\begin{equation*}
\mathrm{A}_{-}(\mathrm{d}{\nu })\mathrm{A}_{+}{(\mathrm{d}\nu }^{\prime }{)}=%
{\delta }_{\nu {\nu }^{^{\prime }}}d\nu ,\ \ \ \mathrm{A}^{+}(\mathrm{d}{\nu 
})\mathrm{A}^{-}{(\mathrm{d}\nu }^{\prime }{)}=0,
\end{equation*}%
\begin{equation*}
\mathrm{A}^{+}(\mathrm{d}{\nu })\mathrm{A}_{+}{(\mathrm{d}\nu }^{\prime }{)}%
=0,\ \ \ \mathrm{A}_{-}(\mathrm{d}{\nu })\mathrm{A}^{-}{(\mathrm{d}\nu }%
^{\prime }{)}=0,
\end{equation*}%
for all $\nu ,{\nu }^{^{\prime }}\in \Omega $. This table includes also $%
\mathrm{A}^{-}(\mathrm{d}{\nu })\mathrm{A}^{+}{(\mathrm{d}\nu }^{\prime }{)}=%
{\delta }_{\nu {\nu }^{^{\prime }}}d\nu $ and all other products equal zero
by the reflection $\nu \mapsto -\nu $.

Let us prove that an arbitrary (not necessary vacuum) stationary (in the
second order sense) quantum process $\mathrm{y}\left( t\right) $ together
with its time-reversed version $\widetilde{\mathrm{y}}\left( t\right) $ can
be obtained by quantum stochastic integration with respect to the
flip-selfadjoint canonical pair $(\mathrm{A}^{+},\mathrm{A}_{-})$ over $%
\Omega \subseteq \mathbf{R}$, or, equivalently, with respect to the
self-adjoint quadruple $\left( \mathrm{A}^{+},\mathrm{A}^{-};\mathrm{A}_{-},%
\mathrm{A}_{+}\right) $ on the positive part $\Omega _{+}\subseteq \mathbf{R}%
_{+}$ of $\Omega $.

Indeed, in general the pair%
\begin{equation*}
\mathrm{y}(t)=\int_{{}}^{{}}\exp \left\{ {-2\pi \mathrm{i}\nu t}\right\} 
\mathrm{\check{Y}}(\mathrm{d}\nu ),\ \ \ \widetilde{\mathrm{y}}%
(t)=\int_{{}}^{{}}\exp \left\{ {-2\pi \mathrm{i}\nu t}\right\} \mathrm{\hat{Y%
}}(\mathrm{d}\nu )
\end{equation*}%
of operator-valued distributions $\check{c}\mapsto \check{c}\cdot \mathrm{y},%
\check{c}\cdot \widetilde{\mathrm{y}}$ is given by a pair $(\mathrm{\check{Y}%
},\mathrm{\hat{Y}})$ of operator-valued orthogonal flip-selfadjoint measures 
$\mathrm{\hat{Y}}\left( \mathrm{d}\nu \right) ^{\dagger }=\mathrm{\check{Y}}%
\left( -\mathrm{d}\nu \right) $ with multiplication table 
\begin{equation*}
\mathrm{\check{Y}}(\mathrm{d}\nu )\mathrm{\check{Y}}{(\mathrm{d}\nu )}^{\dag
}=\sigma {(\nu )}^{2}\mathrm{d}\nu ,\ \ \ \mathrm{\hat{Y}}(\mathrm{d}\nu )%
\mathrm{\check{Y}}{(\mathrm{d}\nu )}^{\dag }=\widetilde{\sigma }(\nu )\sigma
(\nu )\mathrm{d}\nu ,
\end{equation*}%
\begin{equation*}
\mathrm{\check{Y}}(\mathrm{d}\nu )\mathrm{\hat{Y}}{(\mathrm{d}\nu )}^{\dag
}=\sigma (\nu )\widetilde{\sigma }(\nu )\mathrm{d}\nu ,\ \ \ \mathrm{\hat{Y}}%
(\mathrm{d}\nu )\mathrm{\hat{Y}}{(\mathrm{d}\nu )}^{\dag }=\widetilde{\sigma 
}{(\nu )}^{2}\mathrm{d}\nu .
\end{equation*}%
Such measures can be obtained as quantum stochastic integrals 
\begin{equation*}
\mathrm{\check{Y}}(\Delta )=\int_{\Delta }^{{}}\sigma (\nu )\mathrm{A}_{-}(%
\mathrm{d}\nu )+\widetilde{\sigma }(\nu )\mathrm{A}^{+}(\mathrm{d}\nu )
\end{equation*}%
\begin{equation*}
\mathrm{\hat{Y}}(\Delta )=\int_{\Delta }^{{}}\widetilde{\sigma }(\nu )%
\mathrm{A}_{-}(\mathrm{d}\nu )+\sigma (\nu )\mathrm{A}^{+}(\mathrm{d}\nu )
\end{equation*}%
from the canonical pair $(\mathrm{A}^{+},\mathrm{A}_{-})$ of flip-adjoint
annihilation and creation measures. Moreover, in the nonclassical case $%
\sigma (\nu )\neq \widetilde{\sigma }(\nu ),\forall \nu \in \Omega $,
corresponding to quantum noise of zero or finite temperature, the canonical
pair $\mathrm{A}=(\mathrm{A}^{+},\mathrm{A}_{-})$ is uniquely defined over $%
\Omega $ by the pair $(\mathrm{\check{Y}},\mathrm{\hat{Y}})$ from 
\begin{equation*}
\left( {\widetilde{\sigma }{(\nu )}^{2}-\sigma {(\nu )}^{2}}\right) \mathrm{A%
}^{+}(d\nu )=\widetilde{\sigma }(\nu )\mathrm{\check{Y}}(\mathrm{d}\nu
)-\sigma (\nu )\mathrm{\hat{Y}}(\mathrm{d}\nu ),
\end{equation*}%
\begin{equation*}
\left( {\widetilde{\sigma }{(\nu )}^{2}-\sigma {(\nu )}^{2}}\right) \mathrm{A%
}_{-}(\mathrm{d}\nu )=\widetilde{\sigma }(\nu )\mathrm{\hat{Y}}(\mathrm{d}%
\nu )-\sigma (\nu )\mathrm{\check{Y}}(\mathrm{d}\nu ).
\end{equation*}%
This means that the direct and reversed quantum stochastic integrals $\zeta
\cdot \mathrm{y},\zeta \cdot \widetilde{\mathrm{y}}$ can be written for the
Fourier integrals of $\zeta =\check{c}$ as 
\begin{equation*}
\zeta \cdot \mathrm{y}=\int_{{}}^{{}}{g}\left( \nu \right) \mathrm{A}_{-}(%
\mathrm{d}\nu )+{\tilde{g}}\left( \nu \right) \mathrm{A}^{+}(\mathrm{d}\nu )=%
\mathrm{A}(\sigma {c},\widetilde{\sigma }{c})
\end{equation*}%
with ${g}\left( \nu \right) =\sigma (\nu )c(\nu )$ and 
\begin{equation*}
\zeta \cdot \widetilde{\mathrm{y}}=\int_{{}}^{{}}{\tilde{g}}\left( \nu
\right) \mathrm{A}_{-}(\mathrm{d}\nu )+{g}\left( \nu \right) \mathrm{A}^{+}(%
\mathrm{d}\nu )=\mathrm{A}(\,\widetilde{\sigma }{c},\sigma {c})
\end{equation*}%
with ${\tilde{g}}\left( \nu \right) =\widetilde{\sigma }(\nu )c(\nu )$.

To obtain these integrals in the time representation as linear stationary
filters of the adjoint pair $\mathrm{\hat{a}}=(\mathrm{\hat{a}}^{+},\mathrm{%
\hat{a}}${$_{-}$}$)$ of the canonical annihilation $\mathrm{\hat{a}}%
_{-}\left( t\right) $ and creation $\mathrm{\hat{a}}^{+}\left( t\right) $
distributions 
\begin{eqnarray*}
\mathrm{\hat{a}}_{-}\left( t\right) &=&\int_{{}}^{{}}\exp \left\{ -{2\pi 
\mathrm{i}\nu t}\right\} \mathrm{A}_{-}(\mathrm{d}\nu )=\mathrm{\hat{a}}%
^{+}\left( t\right) ^{\dagger },\qquad \\
\mathrm{\hat{a}}^{+}\left( t\right) &=&\int_{{}}^{{}}\exp \left\{ -{2\pi 
\mathrm{i}\nu t}\right\} \mathrm{A}^{+}(\mathrm{d}\nu )=\mathrm{\hat{a}}%
_{-}\left( t\right) ^{\dagger },
\end{eqnarray*}%
or their time reversals $\mathrm{\check{a}}^{-}\left( t\right) =\mathrm{\hat{%
a}}_{-}\left( -t\right) $, $\mathrm{\check{a}}_{+}\left( t\right) =\mathrm{%
\hat{a}}^{+}\left( -t\right) $ we can use the Fourier-Parseval identity 
\begin{equation*}
\mathrm{A}({\widehat{\varphi }}^{-},{\widehat{\varphi }}_{+})=\mathrm{\hat{a}%
}({\varphi }^{-},{\varphi }_{+}).
\end{equation*}%
The latter is given by the standard creation and annihilation integrators 
\begin{equation*}
\mathrm{\hat{A}}^{+}(\mathrm{d}t)=\mathrm{\hat{a}}^{+}\left( t\right) 
\mathrm{d}t=\ \mathrm{\hat{A}}_{-}(\mathrm{d}t)^{\dagger },\ \ \ \mathrm{%
\hat{A}}_{-}(\mathrm{d}t)=\mathrm{\hat{a}}_{-}\left( t\right) \mathrm{d}t=%
\mathrm{\hat{A}}^{+}\left( \mathrm{d}t\right) ^{\dagger }
\end{equation*}%
in the time representation as the standard quantum stochastic integral 
\begin{equation*}
\mathrm{\hat{a}}({\varphi }^{-},{\varphi }_{+})=\int_{-\infty }^{\infty }{%
\varphi }_{+}\left( t\right) \mathrm{\hat{A}}^{+}(\mathrm{d}t)+{\varphi }%
^{-}\left( t\right) \mathrm{\hat{A}}_{-}(\mathrm{d}t)
\end{equation*}%
Applying this to the quantum stochastic integral in the spectral domain with
the general integrands 
\begin{equation*}
{f}^{-}\left( \nu \right) =a(\nu )\widetilde{\sigma }(\nu )+c(\nu )\sigma
(\nu )=\widehat{{\varphi }}^{-}\left( \nu \right)
\end{equation*}%
\begin{equation*}
{f}_{+}\left( \nu \right) =a(\nu )\sigma (\nu )+c(\nu )\widetilde{\sigma }%
(\nu )=\widehat{{\varphi }}_{+}\left( \nu \right)
\end{equation*}%
corresponding to $\check{a}\cdot \mathrm{y}+\check{c}\cdot \widetilde{%
\mathrm{y}}=\mathrm{A}({f}^{-},{f}_{+})$, we obtain $\check{a}\cdot \mathrm{y%
}+\check{c}\cdot \widetilde{\mathrm{y}}=\mathrm{\hat{a}}({\ \varphi }^{-},{%
\varphi }_{+})$, where the components 
\begin{equation*}
{\varphi }^{-}\left( t\right) =(\check{\sigma}\ast \widetilde{\xi })(-t)+(%
\check{\sigma}\ast \eta )(t)={\widehat{f}}^{-}\left( t\right)
\end{equation*}%
\begin{equation*}
{\varphi }_{+}\left( t\right) =(\check{\sigma}\ast \xi )(t)+(\hat{\sigma}%
\ast \eta )(-t)={\widehat{f}}_{+}\left( t\right)
\end{equation*}%
are given by the convolutions 
\begin{equation*}
(\check{\sigma}\ast \zeta )\left( t\right) =\int\limits_{-\infty }^{\infty }%
\check{\sigma}\left( t-r\right) \zeta \left( r\right) \mathrm{d}r
\end{equation*}%
with transition function 
\begin{equation*}
\check{\sigma}(t)=\int_{{}}^{{}}\exp \left\{ {2\pi \mathrm{i}\nu t}\right\}
\sigma (\nu )\mathrm{d}\nu =\hat{\sigma}\left( -t\right)
\end{equation*}%
Thus, the generalized processes $\mathrm{y}$ and $\widetilde{\mathrm{y}}$
are obtained by the stationary filters 
\begin{equation*}
\mathrm{y}(t)=\int_{-\infty }^{\infty }\check{\sigma}(s-t)\mathrm{\hat{A}}%
_{-}(\mathrm{d}s)+\int_{-\infty }^{\infty }\hat{\sigma}(s-t)\mathrm{\hat{A}}%
^{+}(\mathrm{d}s)
\end{equation*}%
\begin{equation*}
\widetilde{\mathrm{y}}(t)=\int_{-\infty }^{\infty }\hat{\sigma}(s-t)\mathrm{%
\hat{A}}_{-}(\mathrm{d}s)+\int_{-\infty }^{\infty }\check{\sigma}(s-t)%
\mathrm{\hat{A}}^{+}(\mathrm{d}s)
\end{equation*}%
corresponding to the quantum stochastic integral representation 
\begin{equation*}
\zeta \cdot \mathrm{y}=\int_{-\infty }^{\infty }(\check{\sigma}\ast \zeta
)(s)\mathrm{\hat{A}}_{-}(\mathrm{d}s)+\int_{-\infty }^{\infty }(\hat{\sigma}%
\ast \widetilde{\zeta })(s)\mathrm{\hat{A}}^{+}(\mathrm{d}s),
\end{equation*}%
\begin{equation*}
\zeta \cdot \widetilde{\mathrm{y}}=\int_{-\infty }^{\infty }(\hat{\sigma}%
\ast \widetilde{\zeta })(s)\mathrm{\hat{A}}_{-}(\mathrm{d}s)+\int_{-\infty
}^{\infty }(\check{\sigma}\ast \zeta )(s)\mathrm{\hat{A}}^{+}(\mathrm{d}s)
\end{equation*}%
for the test functions $\widetilde{\zeta }(t)=\zeta (-t)$ with the square
integrable $\widehat{\zeta }\sigma $ and $\widehat{\zeta }\widetilde{\sigma }
$.

\section*{Conclusion}

Thus we proved that for each quantum stationary noise there exists a
fundamental output process which replaces the noise if the time is reversed.
The nondemolition observation of the quantum noise%
\index{quantum noise|)} via the measurement of the fundamental output process%
\index{output process|)} provides the best mean square filtering of the
noise and in the classical limit completely eliminates this noise.

The input and output stationary processes can be decomposed in the second
order into an orthogonal pair, consisting of vacuum and thermal noises which
are orthogonal to the vacuum and connected by a reversible input-output
(modular) filter. The time-continuous representation of the stationary
linear filters requires a second order theory of quantum stochastic
integration.

The direct application of standard quantum stochastic integration with
respect to the quantum white noise integrators is not possible for
equilibrium (KMS) states because of the non-Markovian character of the
corresponding standard quantum processes $\mathrm{x}$ and $%
\widetilde{\mathrm{x}}$.

But it is possible to modify this approach for the spectral domain where the
stationary quantum processes are $\delta $-correlated. Using this approach
in the second order, we proved that they can be canonically decomposed into
the superpositions of the vacuum pair of direct and reversed noises and can
be obtained by a linear filtering%
\index{linear filtering} from the standard vacuum noises $\mathrm{x}$ and $%
\widetilde{\mathrm{x}}$.

In general the spectral quantum stochastic integrates need not to be adapted
and homogeneous with respect to the frequency shifts and the treatment of
quantum nonlinear filters and other transformations of quantum equilibrium
noise requires non stationary and non adapted theory of quantum stochastic
integration%
\index{stochastic integration|)} \cite{bib:noise8}.

\vspace{0.5cm} \noindent \textbf{Acknowledgment} \vspace{0.2cm} \newline
\noindent The first author (VPB) is grateful for the hospitality and support
of the Tamagawa University where this paper was partially written. 


\end{document}